\documentstyle[11pt,a4]{article}
\def\be{\begin{equation}}
\def\ee{\end{equation}}
\def\beq{\begin{eqnarray}}
\def\eeq{\end{eqnarray}}

\begin{document}

\thispagestyle{empty}

\begin{flushright}
TIFR/TH/95-08
\end{flushright}
\bigskip\bigskip\bigskip

\begin{center}
{\LARGE {\bf CHARGED HIGGS BOSON SEARCH AT THE TEVATRON UPGRADE
USING TAU POLARIZATION}} \\
\bigskip\bigskip

{\large Sreerup Raychaudhuri and D. P. Roy}\footnote{e-mail:
dproy@theory.tifr.res.in} \\
\bigskip\bigskip

{\em Theoretical Physics Group \\ Tata Institute of Fundamental
Research \\ Homi Bhabha Road, Bombay 400 005, India} \\
\bigskip\bigskip

{\large\bf ABSTRACT}
\end{center}
\bigskip\bigskip

We explore the prospect of charged Higgs boson search in top quark decay
at the Tevatron collider upgrade, taking advantage of the opposite states
of $\tau$ polarization resulting from the $H^\pm$ and $W^\pm$ decays.
Methods of distinguishing the two contributions in the inclusive 1-prong
hadronic decay channel of $\tau$ are suggested.  The resulting signature
and discovery limit of $H^\pm$ are presented for the Tevatron upgrade as
well as the Tevatron$^\star$ and the Ditevatron options. \\
\vspace{1in}

\begin{flushleft}
March 1995
\end{flushleft}

\newpage
\section{INTRODUCTION}

There is indirect evidence for the existence of top quark in the mass
region of
\be
m_t \simeq 175 \ {\rm GeV}
\ee
from the precision measurements of electro-weak parameters, particularly
at LEP \cite{review}.  Moreover, a promising top quark signal in this mass
range has been recently observed by the CDF and $D0\!\!\!/$
collaborations \cite{cdf} at
the Tevatron $\bar pp$ collider.  The ongoing Tevatron collider
experiments by the CDF and $D0\!\!\!/$ collaborations are accumulating a
luminosity of $\sim 100$ pb$^{-1}$ each, which is expected to yield a few
tens of top quark events for the mass range of (1).  Thus one expects to
see a more definitive signal of top quark production from these
experiments at the end of this run.  The upgradation of the Tevatron
collider luminosity via the
installation of the main injector following this run is scheduled to give
a typical accumulated luminosity of
\be
\int {\cal L} \ dt \sim 2 \ {\rm fb}^{-1},
\ee
which corresponds to several hundred top quark events for the above
mentioned mass range (1).  This will enable us to search for new particles
in top quark decay; the large top quark mass offers the possibility of
carrying on this search to a hitherto unexplored mass range for these
particles.  There has been a good deal of recent interest in the search
for one such new particle, for which the top quark decay provides by far
the best discovery limit \cite{barger,drees}.  This is the charged Higgs
boson of the two-Higgs doublet models and in particular the minimal
supersymmetric standard model (MSSM).

Generally the charged Higgs signature in top quark decay is based on its
preferential coupling to the $\tau$ channel vis-\`a-vis $e$ and $\mu$, in
contrast to the universal $W$ coupling to all the three channels.  Thus a
departure from the universality prediction between these decay channels
can be used to separate the charged Higgs signal from the $W$ boson
background in
\be
t \rightarrow bH(W) \rightarrow b\tau\nu.
\ee
Moreover the charged Higgs and the $W$ boson decays lead to opposite
states of $\tau$ polarization, {\it i.e.}
\be
H^- \rightarrow \tau^-_R \bar\nu_R, \ H^+ \rightarrow \tau^+_L
\nu_L
\ee
and
\be
W^- \rightarrow \tau^-_L \bar\nu_R, \ W^+ \rightarrow \tau^+_R
\nu_L
\ee
which can be used to augment the above signature (or even as an
independent signature) \cite{barnett,bullock}.  The present work is
devoted to a quantitative analysis of the signature and discovery limit of
the charged Higgs boson at the Tevatron upgrade based on the above ideas.
In particular it shows how the $\tau$ polarization effect can be exploited
to improve the signature and the discovery limit of the charged Higgs
boson even without identifying the mesonic states in $\tau$ decay, which
will be the case at a hadron collider.
\bigskip

\section{CHARGED HIGGS SIGNAL IN TOP QUARK DECAY}

We shall concentrate on the charged Higgs boson of the MSSM.  Its
couplings to fermions are given by
\beq
{\cal L} = \frac{g} {\sqrt{2} m_W} H^+ \Bigg\{\cot \beta \ V_{ij}
m_{u_i} \bar u_{i} d_{jL} &+& \tan \beta \ V_{ij} m_{d_j} \bar u_{i} d_{jR}
\nonumber \\[2mm] &+& \tan \beta \ m_{\ell_j} \bar\nu_{j} \ell_{jR}\Bigg\}
+ H.c.
\eeq
where $V_{ij}$ are the Kobayashi-Maskawa (KM) matrix elements and $\tan
\beta$ is the ratio of the vacuum expectation values of the two Higgs
doublets.  The QCD corrections are taken into account in the leading log
approximation by substituting the quark mass parameters by their running
masses evaluated at the $H^\pm$ mass scale \cite{drees}.  Perturbative
limits on the $tbH$ Yukawa couplings of Eq. (6), along with the
constraints from the low energy processes like $b \rightarrow s\gamma$ and
$B_d-\bar B_d$ mixing, imply the limits \cite{barger2}
\be
0.4 < \tan\beta < 120.
\ee
In the most predictive form of MSSM, characterised by a common SUSY
breaking mass term at the grand unification point, one gets stronger
limits \cite{ridolfi}
\be
1 < \tan\beta < m_t/m_b.
\ee
Such a lower bound also follows from requiring the perturbative limit on
the $tbH$ Yukawa coupling to hold upto the unification point
\cite{bagger}.

In the diagonal KM matrix approximation, one gets the decay widths

\newpage

\beq
\Gamma_{t \rightarrow bW} &=& \frac{g^2}{64\pi m^2_W m_t}
\lambda^{\frac{1}{2}}
\left(1,\frac{m^2_b}{m^2_t},\frac{m^2_W}{m^2_t}\right)\nonumber
\\[2mm] & &
\left[m^2_W (m^2_t + m^2_b) + (m^2_t - m^2_b)^2 - 2m^4_W\right] \\
\Gamma_{t\rightarrow bH} &=& \frac{g^2}{64\pi m^2_W m_t}
\lambda^{\frac{1}{2}}
\left(1,\frac{m^2_b}{m^2_t},\frac{m^2_H}{m^2_t}\right) \nonumber \\
[2mm] & & \left[(m^2_t \cot^2\beta + m^2_b \tan^2\beta) (m^2_t + m^2_b -
m^2_H) - 4m^2_t m^2_b\right] \\
\Gamma_{H \rightarrow \tau\nu} &=& \frac{g^2m_H}{32\pi m^2_W} m^2_\tau
\tan^2 \beta \\
\Gamma_{H \rightarrow c\bar s} &=& \frac{3g^2 m_H}{32\pi m^2_W}
\left(m^2_c \cot^2 \beta + m^2_s \tan^2 \beta\right).
\eeq
\noindent From these one can construct the relevant branching fractions
\be
B_{t \rightarrow bH} = \Gamma_{t \rightarrow bH}\big/\left(\Gamma_{t
\rightarrow bH} + \Gamma_{t \rightarrow bW}\right)
\ee
\be
B_{H \rightarrow \tau \nu} = \Gamma_{H \rightarrow
\tau\nu}\big/\left(\Gamma_{H \rightarrow \tau\nu} + \Gamma_{H
\rightarrow c\bar s}\right).
\ee
It is the product of these two branching fractions that controls the size
of the observable charged Higgs signal.  The $t \rightarrow bH$ branching
fraction has a pronounced dip at
\be
\tan\beta = (m_t/m_b)^{\frac{1}{2}} \simeq 6,
\ee
where (10) has a minimum.  Although this is partly compensated by a large
value of the $H \rightarrow \tau\nu$ branching fraction, which is $\simeq
1$ for $\tan\beta > 2$, the product still has a significant dip at (15).
Consequently the predicted charged Higgs signal will be very weak around
this point as we shall see below.

The basic process of interest is $t\bar t$ pair production through
gluon-gluon (or quark-antiquark) fusion followed by their decay into
charged Higgs or $W$ boson channels, i.e.
\be
gg \rightarrow t\bar t \rightarrow b\bar b (H^+H^-, H^\pm W^\mp, W^+ W^-).
\ee
The $\tau$ decay (4,5) of one or both the charged bosons leads to a single
$\tau$, $\tau\tau$ or $\ell\tau$ final state, where $\ell$ denotes $e$ and
$\mu$.  Each of these final states is accompanied by a large missing-$E_T$
and several hadronic jets.

A brief discussion of the $\tau$-identification at hadron colliders is in
order here. Starting with a missing-$E_T$ trigger, the UA1, UA2 and CDF
experiments have been able to identify $\tau$ as a narrow jet in its
hadronic decay mode \cite{ua2,cdf2}.  In particular the CDF experiment has
used the narrow jet cut to reduce the QCD jet background by an order of
magnitude while retaining most of the hadronic $\tau$ events.  Moreover,
since the hadronic $\tau$ and QCD jet events dominantly populate the
1-prong and multi-prong channels respectively, the prong distribution of
the narrow jets can be used to distinguish the two.  This way the CDF
experiment \cite{cdf2} has been able to identify the $W \rightarrow
\tau\nu$ events and test $W$ universality as well as put modest
constraints on $t$ and $H^\pm$ masses from (3) using a data sample of
integrated luminosity $\sim 4 $ pb$^{-1}$.  In the present case, however,
one would be looking for a few tens of hadronic $\tau$ events in a data
sample of $\sim 500$ times larger integrated luminosity, for which the QCD
jet background cannot be controlled by the above method.  Therefore one
cannot use the single $\tau$ channel for the charged Higgs search and even
the $\tau\tau$ channel may be at best marginal.  The best charged Higgs
signature is provided by the $\ell\tau$ channel.  The largest background
comes from $W \rightarrow \ell\nu$ accompanied by QCD jets, which can be
easily suppressed by the above mentioned jet angle and multiplicity cuts.
Besides the hard isolated lepton $\ell$ provides a more robust trigger
than the missing-$E_T$.  Therefore in this work we shall concentrate
mainly on the $\ell\tau$ channel; but similar analysis can be carried over
in the $\tau\tau$ channel as well.

The $\ell\tau$ and $\tau\tau$ channels correspond to the leptonic decay of
both the charged bosons in (16), i.e.
\beq
& H^+ & \ H^- \ \ , \ \ H^+ \ \ \ W^- \ \ , \ \ H^- \ \ \ W^+ \ \ ,
\ \ \ W^+ \ \ \ \ \ \ \ W^- \nonumber \\[2mm]
& \downarrow & \ \ \downarrow \ \ \ \ \ \ \ \
\downarrow \ \ \ \ \ \ \downarrow \ \ \ \ \ \ \ \ \
\downarrow \ \ \ \ \ \ \downarrow \ \ \ \ \ \ \ \ \ \ \
\downarrow \ \ \ \ \ \ \ \ \ \ \downarrow \nonumber \\[2mm]
& \tau^+_L & \ \tau^-_R \ \ \ \ \ \tau^+_L \ \ \ \tau^-_L,\ell^- \ \ \ \ \
\tau^-_R \ \ \ \tau^+_R,\ell^+ \ \ \ \ \ \tau^+_R,\ell^+ \ \ \
\tau^-_L,\ell^-
\eeq
By convention,
\be
P_\tau \equiv P_{\tau^-} = -P_{\tau^+}, \ \ \ P_{\tau^\pm} =
\frac{\sigma_{\tau^\pm_R} - \sigma_{\tau^\pm_L}}{\sigma_{\tau^\pm_R} +
\sigma_{\tau^\pm_L}}.
\ee
For the $\ell\tau$ channel of our interest the signal and the background
come from the $HW$ and $WW$ terms respectively.  They correspond to
exactly opposite states of $\tau$ polarization, {\em i.e.}
\be
P^H_\tau = + 1, \ \ \ P^W_\tau = -1.
\ee
Consequently the use of the tau polarization effect for enhancing the
signal to background ratio is particularly simple in this case as we shall
see below.  It may be noted here that the $\tau\tau$ channel has a better
signal to background ratio because of the $HH$ contribution as well as the
enhancement of $WH$ relative to $WW$ by a combinatorial factor of 2.  On
the other hand the polarisation distinction is less clean. While both the
$\tau$'s in the background have negative polarisation one or both of them
have positive polarisation in the signal.  Nonetheless the method of
enhancing the signal to background ratio by the $\tau$ polarization effect
discussed below can be extended to this channel, provided one can identify
the $\tau\tau$ events from the QCD background.
\bigskip

\section{TAU POLARIZATION EFFECT}

We shall concentrate on the 1-prong hadronic decay channel of $\tau$,
which is best suited for $\tau$ identification.  It accounts for 80\% of
hadronic $\tau$ decays and 50\% of overall $\tau$ decays.  The main
contributors to the 1-prong hadronic $\tau$ decay are \cite{review}
\beq
\tau^\pm & \rightarrow & \pi^\pm \nu_\tau  \ (12.5\%)  \\
\tau^\pm & \rightarrow & \rho^\pm \nu_\tau \rightarrow \pi^\pm \pi^0 \nu_\tau
\ (24\%)  \\
\tau^\pm & \rightarrow & a^\pm_1 \nu \rightarrow \pi^\pm \pi^0 \pi^0 \nu_\tau
 \ (7.5\%)
\eeq
where the branching fractions for the $\pi$ and $\rho$ channels include
the small contributions from the $K$ and $K^\star$ channels respectively,
since they have identical polarization effects.  Note that only half the
$a_1$ decay channel contributes to the 1-prong configuration.  The masses
and widths of $\rho$ and $a_1$ are
\cite{review}
\be
m_\rho (\Gamma_\rho) = 770 (150) \ {\rm MeV}, \ \ \ m_{a_1} (\Gamma_{a_1})
= 1260 (400) \ {\rm MeV}.
\ee
One sees that the three decay processes (20,21,22) account for about 90\%
of the 1-prong hadronic decay of $\tau$.  Thus the inclusion of $\tau$
polarization effect in these processes will account for its effect in the
inclusive 1-prong hadronic decay channel to a good approximation.

The formalism relating $\tau$ polarization to the momentum distribution of
its decay particles in (20,21,22) has been widely discussed in the
literature \cite{barnett,bullock,tsai,rouge}.  We shall only discuss the
main formulae relevant for our analysis.  A more detailed account can be
found in a recent paper by Bullock, Hagiwara and Martin \cite{bullock},
which we shall closely follow.  For $\tau$ decay into $\pi$ or a vector
meson $(\rho,a_1)$, one has
\beq
\frac{1}{\Gamma_\pi} \frac{d\Gamma_\pi} {d \cos\theta}
&=& \frac{1}{2} (1 + P_\tau \cos \theta)
\\
\frac{1}{\Gamma_v} \frac{d\Gamma_{vL}}{d\cos\theta}
&=& \frac{\frac{1}{2} m^2_\tau}{m^2_\tau + 2m^2_v} (1 + P_\tau \cos\theta)
\\
\frac{1}{\Gamma_v} \frac{d\Gamma_{vT}}{d \cos\theta}
&=& \frac{m^2_v}{m^2_\tau + 2m^2_v} (1 - P_\tau \cos\theta)
\eeq
where $v$ stands for the vector meson and $L,T$ denote its longitudinal
and transverse polarization states.  The angle $\theta$ measures the
direction of the meson in the $\tau$ rest frame relative to the $\tau$
line of flight, which defines its polarization axis.  It is related to the
fraction $x$ of the $\tau$ energy-momentum carried by the meson in the
laboratory frame, {\it i.e.}
\be
\cos\theta = \frac{2x - 1 - m^2_{\pi,v}/m^2_\tau}{1 - m^2_{\pi,v}/m^2_\tau}.
\ee
Here we have made the collinear approximation $m_\tau \ll p_\tau$, where
all the decay products emerge along the $\tau$ line of flight in the
laboratory frame.

The above distribution (24-26) can be simply understood in terms of
angular momentum conservation.  For $\tau^-_{R(L)} \rightarrow \nu_L \
\pi^-$, $v^-_{\lambda=0}$ it favours forward (backward) emission of $\pi$
or longitudinal vector meson, while it is the other way round for
transverse vector meson emission $\tau^-_{R(L)} \rightarrow \nu_L
v^-_{\lambda=-1}$.  Thus the $\pi^\pm$s coming from $H^\pm$ and $W^\pm$
decays peak at $x=1$ and $0$ respectively and $\langle x_\pi\rangle_H =
2\langle x_\pi\rangle_W = 2/3$.  Although the clear separation between the
signal and the background peaks disappears after convolution with the
$\tau$ momentum, the relative size of the average $\pi$ momenta remains
unaffected, {\em i.e.}
\be
\langle p^T_\pi \rangle_H \simeq 2 \langle p^T_\pi \rangle_W \ \ {\rm
for} \ \ m_H \simeq m_W.
\ee
Thus the $\tau$ polarization effect (24) is reflected in a significantly
harder $\pi^\pm$ momentum distribution for the charged Higgs signal
compared to the $W$ boson background.  The same is true for the
longitudinal vector mesons; but the presence of the transverse component
dilutes the polarization effect in the vector meson momentum distribution
by a factor (see eqs. 25,26)
\be
\frac{m^2_\tau - 2m^2_v}{m^2_\tau + 2m^2_v}.
\ee
Consequently the effect of $\tau$ polarization is reduced by a factor of
$\sim 1/2$ in $\rho$ momentum distribution and practically washed out in
the case of $a_1$.  Thus the inclusive 1-prong $\tau$ jet resulting from
(20-22) is expected to be harder for the $H^\pm$ signal compared to the
$W$ boson background; but the presence of the transverse $\rho$ and $a_1$
contributions makes the size of this difference rather modest.  We shall
see below that it is possible to suppress the transverse $\rho$ and $a_1$
contributions and enhance the difference between the signal and the
background in the 1-prong hadronic $\tau$ channel even without identifying
the individual mesonic contributions to this channel.

The key feature of vector meson decay, relevant for the above purpose, is
the correlation between its state of polarization and the energy sharing
among the decay pions.  In order to use this feature, one must first
transform the polarization states of the vector meson appearing in (25,26)
from the $\tau$ rest frame to the laboratory frame.  This is done by a
Wigner rotation of the vector meson spin quantization axis \cite{martin},
i.e.
\be
M'_{\lambda_\tau \lambda'_v} = \sum_{\lambda_v} d^1_{\lambda'_v
\lambda_v} (\omega) M_{\lambda_\tau \lambda_v}
\ee
where the decay helicity amplitudes on the left and right correspond to
the laboratory and the $\tau$ rest frames respectively; and
\be
\cos \omega = \frac{(m^2_\tau - m^2_v) + (m^2_\tau + m^2_v)
\cos\theta} {(m^2_\tau + m^2_v) + (m^2_\tau - m^2_v) \cos \theta}
\ee
in the collinear approximation.  It may be noted that over most of the
range of $\cos\theta$ the angle $\omega$ remains very small for $\rho$ and
to a lesser extent for $a_1$ as well.  Thus the longitudinal and
transverse states of the vector meson polarization in the $\tau$ rest
frame roughly correspond to those in the laboratory frame, so that the
suppression of the transverse state in the latter frame corresponds to its
suppression in the former as well.  Using (30,31) one can rewrite the
decay formulae (25) and (26) in terms of the polarization states in the
laboratory frame, i.e.
\beq
\frac{1}{\Gamma_v} \frac{d\Gamma^{\rm lab}_{vL}}{d \cos\theta}
&=& \frac{\frac{1}{2} m^2_v} {m^2_\tau + 2m^2_v} \Bigg[\sin^2 \omega +
\frac{m^2_\tau}{m^2_v} \cos^2 \omega + P_\tau \cos \theta \nonumber \\[2mm]
& &  \left(\frac{m_\tau}{m_v} \sin 2\omega \tan \theta +
\frac{m^2_\tau}{m^2_v} \cos^2 \omega - \sin^2 \omega\right)\Bigg]
\\
\frac{1}{\Gamma_v} \frac{d\Gamma^{\rm lab}_{vT}}{d \cos\theta}
&=& \frac{\frac{1}{2} m^2_v}{m^2_\tau + 2m^2_v} \Bigg[1 +\cos^2\omega +
\frac{m^2_\tau}{m^2_v} \sin^2 \omega + P_\tau \cos\theta \nonumber \\[2mm]
& & \left(\frac{m^2_\tau}{m^2_v} \sin^2 \omega -
\frac{m_\tau}{m_v} \sin 2\omega \tan\theta - \cos^2 \omega -
1\right)\Bigg].
\eeq
To take account of the width of the vector meson, (32) and (33) are
averaged over the vector resonance shape function \cite{bullock}.
\be
F_v (m^2) = \left(1 - \frac{m^2}{m^2_\tau}\right)^2 \left(1 +
\frac{2m^2}{m^2_\tau}\right) |D_v(m^2)|^2  f_v (m^2)
\ee
where
\be
D_v (m^2) = \frac{1}{m^2 - m^2_v + im\Gamma_v (m^2)}
\ee
is the vector meson propagator with invariant mass $m^2$ and the running
width
\be
\Gamma_v (m^2) = \Gamma_v \frac{m}{m_v} \frac{f_v(m^2)}{f_v(m^2_v)}.
\ee
The $\rho$ meson line shape factor is
\be
f_\rho (m^2) = (1 - 4m^2_\pi/m^2)^{3/2}
\ee
which takes account of the $P$-wave threshold behavior for $\rho
\rightarrow \pi\pi$ decay.  For the line shape of the $a_1$ meson we
shall use the phenomenological parametrisation of Kuhn and Santamaria
\cite{kuhn}
\be
f_a (m^2) = 1.623 + \frac{10.38}{m^2} - \frac{9.32}{m^4} +
\frac{0.65}{m^6}.
\ee
The $\rho^\pm \rightarrow \pi^\pm \pi^0$ decay distributions for the two
polarization states of (32) and (33) are given by
\beq
\frac{1}{\Gamma_{\pi\pi}}
\frac{d\Gamma (\rho^\pm_L \rightarrow \pi^\pm \pi^0)}{d\cos\theta'}
&=& \frac{3}{2} \cos^2 \theta' \simeq \frac{3}{2} (2x' - 1)^2 \\
\frac{1}{\Gamma_{\pi\pi}}
\frac{d\Gamma (\rho^\pm_T \rightarrow \pi^\pm \pi^0)}{d\cos \theta'}
&=& \frac{3}{4} \sin^2 \theta' \simeq 3x' (1 - x') \\ x^\prime &=& \left[1
+ \sqrt{1 - 4m^2_\pi/m^2_\rho} \cos \theta'\right]/2
\eeq
where $\theta'$ is the angle of the pion pair in the $\rho$ rest frame
measured relative to the $\rho$ line of flight in the laboratory frame,
and $x'$ is the fraction of the $\rho$ energy-momentum carried by one of
the pions (the charged one, say) in the laboratory frame.  Thus $\rho_T$
decay favours equipartition of its laboratory energy between the two
pions, while $\rho_L$ decay favours the asymmetric configurations where
one of the pions carries all or none of its energy.

The $a_1$ decays dominantly via $\rho$, {\em i.e.}
\be
a^\pm_1 \rightarrow \rho^\pm \pi^0 \rightarrow \pi^\pm_1 \pi^0_2
\pi^0_3
\ee
gives the 1-prong decay of our interest. However, one cannot, in general,
predict the energy distribution among the three pions coming from $a_{1L}$
or $a_{1T}$ decay, since each will contain $\rho_L$ and $\rho_T$
contributions with unknown relative strength.  So one has to assume a
dynamical model.  We shall follow the model of Kuhn and Santamaria, based
on the chiral limit (conserved axial vector current approximation), which
provides a good description of the $a_1
\rightarrow 3\pi$ data \cite{kuhn}.  In this model the decay amplitude
is given by
\beq
M &=& \epsilon_\nu^{T,L} J^\nu (a^\pm_1 \rightarrow \pi^0_1 \pi^0_2
\pi^\pm_3) \\
J^\nu &\sim& D_{a_1} (s) \left[D_\rho (s_{13}) (\tilde p_3 - \tilde
p_1)^\nu + D_\rho (s_{23}) (\tilde p_3 - \tilde p_2)^\nu\right] \\
\tilde p^\nu_i &=& p^\nu_i - p^\nu_{a_1} \frac{p_{a_1}.p_i}{s}
\eeq
where we have dropped a constant multiplicative factor in (44), which will
not be relevant for our analysis.  It will be adequate for our purpose to
evaluate the decay amplitudes (43-45) neglecting the $a_1$ and $\rho$
widths.  The resulting decay distributions for longitudinal and transverse
$a_1$ are given by
\beq
&&\frac{1}{\Gamma_{3\pi}}
\frac{d\Gamma (a_{1L} \rightarrow 3\pi)}{d \cos\theta_a
d \cos\theta_\rho d \phi_\rho} = \frac{\left[\frac{m^2_a + m^2_\rho}{m_a
m_\rho} \cos \theta_a \cos \theta_\rho - 2 \sin \theta_a
\sin \theta_\rho \cos \phi_\rho\right]^2}{\frac{8\pi}{9}
\left[\left(\frac{m^2_a + m^2_\rho}{m_a m_\rho}\right)^2 + 8\right]}
\ \ \ \\
&&\frac{1}{\Gamma_{3\pi}}
\frac{d\Gamma (a_{1T} \rightarrow 3\pi)}{d \cos \theta_a
d \cos\theta_\rho d \phi_\rho} \nonumber \\[2mm] &&  \ \ \ \ \ =
\frac{\left[\frac{m^2_a + m^2_\rho}{m_a m_\rho} \sin \theta_a \cos
\theta_\rho + 2 \cos \theta_a
\sin \theta_\rho \cos \phi_\rho\right]^2 + 4\sin^2 \theta_\rho \sin^2
\phi_\rho} {\frac{16\pi}{9} \left[\left(\frac{m^2_a + m^2_\rho}{m_a
m_\rho}\right)^2 + 8\right]}.
\eeq
In the $a^\pm_1 \rightarrow \rho^\pm \pi^0$ decay $\theta_a$ is the angle
of $\rho$ in the $a_1$ rest frame measured relative to the $a_1$ line of
flight in the laboratory ($z$-axis), while the plane containing these two
vectors defines the $x - z$ plane.  Similarly in the $\rho^\pm \rightarrow
\pi^0$ decay $\theta_\rho$ and $\phi_\rho$ are the polar and azimuthal
angles of the charged pion in the $\rho$ rest frame, measured relative to
the above $\rho$ line of flight ($z'$-axis) and the above plane
respectively.  In terms of these angles, the fraction of $a_1$ laboratory
energy-momentum carried by the charged pion is given by
\beq
x' &=& \frac{E_{\pi^\pm}}{E_{a^\pm_1}} \nonumber \\ &=& \frac{1}{m_a}
\Bigg[\frac{m_\rho}{2} \frac{m^2_a + m^2_\rho}{2m_a m_\rho} +
\frac{m_\rho}{2} \frac{m^2_a - m^2_\rho}{2m_a m_\rho} \cos
\theta_a + q \frac{m^2_a - m^2_\rho}{2m_a m_\rho} \cos \theta_\rho
\nonumber \\[2mm] & & + \ \ \ q \frac{m^2_a + m^2_\rho}{2m_a m_\rho} \cos
\theta_\rho \cos \theta_a - q \sin \theta_\rho \cos \phi_\rho \sin
\theta_a\Bigg],
\eeq
\be
q = \frac{1}{2} \sqrt{m^2_\rho - 4m^2_\pi} \simeq \frac{1}{2} m_\rho.
\ee

One sees from (48) that
\be
x' \simeq 1 \ \ {\rm for} \ \ \cos \theta_a \simeq 1 \ \ {\rm and} \ \
\cos \theta_\rho \simeq 1,
\ee
while
\be
x' \simeq 0 \ \ {\rm for} \ \ \cos \theta_a \simeq -1 \ \ {\rm or} \ \
\cos \theta_\rho \simeq -1.
\ee
The $a_{1L}$ decay distribution (46) has maxima near
\be
\cos \theta_a = \pm 1 \ \ \ {\rm along \ with} \ \ \ \cos \theta_\rho = \pm 1,
\ee
which correspond to collinear decay into $3\pi$ resulting in unequal
distribution of energy.  This is similar to the $\rho^\pm_L
\rightarrow \pi^\pm \pi^0$ decay, except that in the present case
there is a visible peak only at $x' \simeq 0$ but not at $x' \simeq 1$.
The reason is that the latter condition holds only for a tiny region of
the phase space as we see from (50).  The $a_{1T}$ decay distribution (47)
vanishes near the collinear configuration (52).  It has maxima at
\be
\cos \theta_a = 0 \ \ {\rm and} \ \ \cos \theta_\rho = \pm 1 \ \ {\rm
or} \ \ \cos \theta_\rho = 0, \ \ \cos \phi_\rho = 0
\ee
which correspond to the plane of the three decay pions in the $a_1$ rest frame
being normal to its line of flight.  This results in an even sharing of
the $a_1$ energy as in the case of $\rho_T$ decay.  In particular both the
distributions vanish at the extrema $x' = 0$ and $1$ and peak near the
middle, although the $a_{1T}$ peak occurs a little below $x' = 0.5$.
Indeed the shapes of $a_{1L}$ and $a_{1T}$ decay distributions in $x'$ are
qualitatively similar to those of $\rho_L$ and $\rho_T$, except for the
suppression of the $x' \simeq 1$ peak for $a_{1L}$.  A comparison of these
distributions can be found in \cite{bullock}.  There is reason to believe
that the above features of longitudinal and transverse $a_1$ decay are
insensitive to the assumed dynamical model \cite{kuhn}.  Indeed it follows
from general considerations that the $a_{1L(T)} \rightarrow 3\pi$ decay
favours the plane of the 3 pions in the $a_1$ rest frame being coincident
with (normal to) the $a_1$ line of flight \cite{rouge}.  The role of the
model is only to determine the distribution of energy among the 3 pions in
this plane. Moreover as shown in \cite{bullock}, the alternative model of
Isgur {\em et al} \cite{isgur} gives very similar pion energy
distributions as that of \cite{kuhn}.

Thus the transverse $\rho$ and $a_1$ decays favour even sharing of energy
by the charged and neutral pions, while the longitudinal $\rho$ and $a_1$
decays favour extreme configurations where the charged pion carries
practically all or none of the vector meson energy.  This can be exploited
to suppress the former while retaining most of the latter contributions
along with that of the pion (20).  This will in turn enhance the $H^\pm$
signal to $W^\pm$ background ratio in the 1-prong hadronic decay channel
of $\tau$ as we shall see below.
\bigskip

\section{RESULTS AND DISCUSSION}

We shall be interested in the inclusive 1-prong hadronic decay of $\tau$,
which is dominated by the $\pi^\pm,\rho^\pm$ and $a^\pm_1$ contribution
(20,21,22).  It results in a thin 1-prong hadronic jet ($\tau$-jet)
consisting of a charged pion along with 0,1 or 2 $\pi^0$ s respectively.
Since all the
pions emerge in a collinear configuration one can neither measure their
invariant mass nor the number of $\pi^0$ s.  Consequently it is not
possible to identify the mesonic state.  But it is possible to measure the
energy of the charged track as well as the total neutral energy, either by
measuring the momentum of the former in the central detector and the total
energy deposit in the EM and hadron calorimeters or from the showering
profiles in the EM and hadron calorimeters.  Thus one has to develop
a strategy to suppress the transverse vector meson contributions using
these two pieces of information. We shall consider two such strategies
below. In either case a rapidity and transverse energy cut of
\be
|\eta| < 2 \ \ {\rm and} \ \ E_T > 20 \ {\rm GeV}
\ee
with be applied on the $\tau$-jet, where $E_T$ includes the neutral
contribution.  We shall use the recent structure functions of
\cite{martin2} for calculating the $t\bar t$ cross-section.

Firstly, we consider the effect of an isolation cut requiring the neutral
$E_T$ accompanying the charged track within a cone of $\Delta R = (\Delta
\eta^2 + \Delta \phi^2)^{1/2} = 0.2$ to be
\be
E^{ac}_T \equiv E^0_T < 5 \ {\rm GeV}.
\ee
Fig. 1 shows the $E^{ac}_T$ distribution for a $\tau$-jet satisfying (54).
The $\pi,\rho$ and $a_1$ contributions are shown separately for the
$H^\pm$ signal and the $W^\pm$ background, where we have chosen $m_H = 80$
GeV and $\tan \beta = 1$ for illustrative purpose.  The $\bar pp$ CM
energy is taken to be 2 TeV.  Several points are worth noting in this
figure. \\ i) The signal to background ratio for $\pi \ (\sim 4.5)$ is
twice as large as $\rho$ and thrice as large as $a_1$.  This is a
consequence of the $E_T > 20$ GeV cut and the $\tau$ polarization effect
(24-29).\\  ii) The $\rho$ and $a_1$ contributions to the signal
(background) are dominated by the longitudinal (transverse) components.\\
iii) The $\rho^\pm_L \rightarrow \pi^\pm \pi^0$ peak at $x'
\simeq 1$ shows up in the signal at $E^{ac}_T \simeq 0$ while the $x'
\simeq 0$ peak is smeared over the large $E^{ac}_T$ tail.  The absence of
a $E^{ac}_T \simeq 0$ peak in the $a^+_{1L} \rightarrow \pi^\pm \pi^0
\pi^0$ contribution to the signal reflects the absence of a corresponding
peak at $x' \simeq 1$ as remarked earlier, while the $x' \simeq 0$ peak is
smeared over the large $E^{ac}_T$ tail. \\ iv) The $\rho^\pm_T \rightarrow
\pi^\pm \pi^0$ peak at $x' \simeq 0.5$ shows up in the background at
$E^{ac}_T \simeq 15 $ GeV.  The peak in the $a^\pm_{1T} \rightarrow
\pi^\pm \pi^0 \pi^0$ contribution to the background at a somewhat higher
$E^{ac}_T$ reflects the corresponding peak at $x'$ somewhat below 0.5 as
remarked earlier.

As one sees from Fig. 1, the isolation cut (55) on the charged track will
essentially remove all the contributions except for $\pi^\pm$ and a part
of the $\rho^\pm_L \rightarrow \pi^\pm \pi^0$ corresponding to its $x'
\simeq 1$ peak, where the decay $\pi^0$ is very soft.  Consequently the
signal to background ratio is enhanced by a factor of $\sim 2$; but the
signal size goes down by a factor of $\sim 2.5$.  Of course the
enhancement of the signal to background ratio increases further with
increasing $E_T$ cut as we shall see below.  Moreover the isolation cut
has the advantage of suppressing the QCD jet background.  Nonetheless the
factor of 2.5 drop in the signal size is a high price to pay, particularly
at the Tevatron collider \cite{iso}.  The reason for this big drop in the
signal size is of course that the isolation cut removes not only the
$\rho_T$ and $a_{1T}$ contributions but also large parts of the $\rho_L$
and $a_{1L}$ contributions corresponding to their $x' \simeq 0$ peaks.
The second strategy discussed below aims at retaining these latter
contributions.

Here one plots the rate of $\tau$-jet events, satisfying (54), as a
function of
\be
\Delta E_T = |E^{ch}_T - E^0_T| = |E^{ch}_T - E^{ac}_T|;
\ee
{\it i.e.} the difference between the $E_T$ of the charged track and the
accompanying neutral $E_T$.  The hard $\tau$-jet events from the $H^\pm$
signal and $W^\pm$ background are expected to be dominated by the
$\pi,\rho_L,a_{1L}$ and the $\rho_T,a_{1T}$ contributions respectively.
The latter contributions favour comparable values of $E^{ch}_T$ and
$E^0_T$ and hence relatively small $\Delta E_T$, while the former favour
large values of $\Delta E_T$.  Thus the signal events are expected to show
significantly harder $\Delta E_T$ distribution compared to the background.

Fig. 2 shows the $\tau$-jet cross-sections from the $H^\pm$ signal and the
$W^\pm$ background for $\tan \beta = 1.4$ and two values of $H^\pm$ mass,
{\it viz.} 100 and 140 GeV.

Fig. 2a and b show the $E_T$ distributions of the inclusive 1-prong $\tau$
jet events from (20,21,22) before and after the isolation cut (55).  The
isolation cut is clearly seen to enhance the signal to background ratio,
but at the cost of a drop in the signal size.  The signal to background
ratio improves by a factor of 1.5 --- 3 over the $E_T$ range shown, while
the signal size drops by a factor of 2-3.  Fig. 2c shows these inclusive
1-prong $\tau$-jet events as a function of $\Delta E_T$.  Evidently the
signal events have a much harder $\Delta E_T$ distribution than the
background, which is far more striking than the difference in the
corresponding $E_T$ distributions shown in Fig. 1a.  Thus the $\Delta E_T$
distribution provides a much clearer separation between the signal and the
background than the simple $E_T$ distribution.  It helps to improve the
signal to background ratio significantly without sacrificing the signal
size.

Fig. 3 shows the corresponding integrated cross-sections against the
cutoff value of the $E_T (\Delta E_T)$, {\it i.e.}
\be
\sigma (E_T) = \int^{E_T}_\infty \frac{d\sigma}{dE_T} dE_T.
\ee
These plots are well suited for comparing the relative merits of the three
methods in extracting the signal from the background.  For this purpose
the cutoff values are to be so chosen that one gets a viable
\be
H^\pm \ {\rm signal}/W^\pm \ {\rm background} \geq 1.
\ee
The resulting signal size is a reasonable criterion for the merit of the
method.  Comparing the signal and background cross-sections for $m_H = 140
$ GeV, we see that this condition is achieved at a far greater sacrifice
to the signal size in Fig. 3a than in $b$ and $c$.  The size of the
resulting signals, as given by the corresponding cross-over points, are
$\sim 1/2$ fb, 3 fb and 7 fb respectively.  Making a similar comparison of
the signal and background cross-sections for $m_H = 100 $ GeV, one sees
that the ratio 1 is reached in 3a with a signal size of $\sim 2$ fb,
which is larger than that in 3b and comparable to the one in 3c.
However, the ratio increases more rapidly with cutoff in the latter
two cases compared
to the first.  Since this increase is required to offset the rapid fall of
the signal to background ratio with increasing $\tan\beta$ (see eqs.
9,10), the latter methods give more favourable signal size at $\tan\beta >
1.4$ as shown in Fig. 4.

Fig. 4a,b,c show the size of the signals from the three methods satisfying
a viable signal to background ratio $ > 1$.  The signal cross-sections are
shown as functions of $\tan\beta$ for $m_H = 80,100,120$ and $140$ GeV.
One clearly sees that the use of $\tau$ polarization effect via the
isolation cut (Fig. 4b) or the $\Delta E_T$ distribution (Fig. 4c) will
give a viable charged Higgs signal over a wider range of the charged Higgs
mass and $\tan\beta$ parameters.

It is reasonable to consider a signal size of 10 fb, satisfying a signal
to background ratio $\geq 1$, to constitute a viable charged Higgs signal.
With the expected integrated luminosity of $\sim 2 $ fb$^{-1}$, this will
correspond to 20 signal events over a $W$ boson background of similar
size.  Since the number of background events can be predicted from the
number of dilepton ($\ell^+ \ell^-$) events in $t\bar t$ decay using $W$
universality, this will correspond to a $4.5
\ \sigma$ signal for the charged Higgs boson.  Thus one can get the
discovery limit of charged Higgs boson at the Tevatron upgrade by
demanding a signal size of 10 fb in Fig. 4.  Evidently the best limits
come from Fig. 4c.  For $m_H = 100 \ (120)$ GeV one expects a viable
signal except for the region $\tan\beta = 2 - 15 \ (1.5-20)$.  The gap in
the $\tan\beta$ space is due to the dip in the $t
\rightarrow bH$ coupling at $\tan\beta \sim 6$, as remarked before.
It may be mentioned here that there is a current suggestion of further
upgradation of Tevatron luminosity by another order of magnitude --- {\em
i.e.} the Tevatron$^\star$.  The corresponding discovery limit of charged
Higgs boson can be obtained by demanding a signal size of 1 fb in Fig.
4c.  In this case the gap narrows down to $\tan\beta = 3 - 10 \ (2.5 -
12)$ for $m_H = 100 \ (120)$ GeV.  Moreover one can probe for $m_H = 140
$ GeV except for a gap in the region $\tan\beta = 2 - 15$.

For the sake of completeness we have computed the signal and background
cross-sections for the suggested Ditevatron energy of $\sqrt{s} = 4$ TeV.
Fig. 5 shows the integrated signal and background cross-sections against
the cutoff $E_T \ (\Delta E_T)$ analogous to Fig. 3 for $m_H = 100$ and
150 GeV.  The curves are very similar to those of Fig. 3 except for a
factor of $\sim 4$ increase in normalisation.  Fig. 6 shows the signal
cross-sections, satisfying signal to background ratio $\geq 1$, as
functions of $\tan\beta$ for $m_H = 80,100,120,140$ and 150 GeV.
Comparing Figs. 4 and 6 one sees better discovery limit at the Ditevatron
for comparable luminosity.  It should be noted, however, that the signal
cross-section of $\sim 10$ fb at the Ditevatron has similar contours in
the $m_H$ and $\tan\beta$ space as that of $\sim 1$ fb at the Tevatron.
Thus one expects similar discovery limits for charged Higgs boson at the
Ditevatron and the Tevatron$^\star$.  In either case there remains a gap
near $\tan\beta \sim 6$, so that the nonobservation of a signal will not
rule out the presence of a charged Higgs boson in the $100 - 140$ GeV
region unambiguously.
\bigskip

\section{SUMMARY}

We have explored the prospect of charged Higgs boson search in top quark
decay at the Tevatron collider upgrade, taking advantage of the opposite
states of $\tau$ polarization resulting from the $H^\pm$ and $W^\pm$
decays.  We have concentrated on the decay of $\tau$ into a 1-prong
hadronic jet ($\tau$-jet), which is dominated by the $\pi^\pm,\rho^\pm$
and $a^\pm_1$ mesons.  The positive (negative) polarisation of $\tau$
coming from the $H^\pm$ signal ($W^\pm$ background) is shown to favour
unequal (equal) sharing of the $\tau$-jet energy between the charged prong
$(\pi^\pm)$ and the accompanying neutral pions.  Consequently the two
polarization states can be distinguished by measuring the charged and
neutral contributions to the 1-prong $\tau$-jet energy even without
identifying the individual meson states.  We have shown how this can be
used for better separation of the charged Higgs signal from the $W$ boson
background.  In particular we have considered two strategies --- 1) an
isolation cut on the $\tau$-jet events requiring the neutral contribution
to the jet transverse energy to be small ($E^0_T < 5 $ GeV), and 2) a
redistribution of the $\tau$-jet events in $\Delta E_T$, {\em i.e.} the
difference between the charged and neutral contributions to the jet $E_T$
instead of their sum.  In either case one gets a substantial enhancement
in the signal to background ratio.  But this is accompalished at the cost
of a reduction in the signal size in the first case, while there is no
such price to pay in the second.  Consequently the latter strategy offers
the best discovery limit for the charged Higgs boson.  We have explored
these discovery limits in the parameter space of $H^\pm$ mass and
$\tan\beta$ assuming an integrated luminosity of $\sim 2 $ fb$^{-1}$ for
the Tevatron upgrade.  For the sake of completeness we have also explored
the signal and discovery limit for the suggested Tevatron$^\star$ and
Ditevatron options, corresponding to an order of magnitude increase of
luminosity and a doubling of the CM energy respectively.
\bigskip

\begin{center}
\underbar{\Large\bf Acknowledgements}
\end{center}

It is a pleasure to thank R. M. Godbole, N. K. Mondal and Probir Roy for
discussions. The work of SR is partially supported by a project (DO No.
SR/SY/P-08/92) of the Department of Science and Technology, Government of
India.

\newpage

\begin{center}
\underbar{\Large\bf Figure Captions}
\end{center}
\bigskip
\begin{enumerate}

\item[{\rm Fig. \ 1.}] The $\pi^\pm, \rho^\pm$ and $a^\pm_1$ contributions
to the 1-prong hadronic $\tau$-jet cross-section coming from the $H^\pm$
signal (upper curves) and $W^\pm$ background (lower curves) for $m_H = 80
$ GeV and $\tan\beta = 1$ at $\sqrt{s} = 2$ TeV.
The cross-sections are shown as functions of
neutral pion $E_T$ accompanying the charged track in the $\tau$-jet.

\item[{\rm Fig. \ 2.}] The 1-prong hadronic $\tau$-jet cross-sections
are plotted against the jet $E_T$ in (a) without and (b) with the
isolation cut.  They are plotted against the $\Delta E_T$ of the jet in
(c).  The $H^\pm$ signal ($W^\pm$ background) contributions are shown as
solid (dashed) lines for $m_H = 100 $ GeV and dot-dashed (dotted) lines
for $m_H = 140 $ GeV.  We take $\sqrt{s} = 2$ TeV and $\tan\beta = 1.4$.

\item[{\rm Fig. \ 3.}]  The integrals of the signal and background
cross-sections of Fig. 2(a,b,c) shown against the cutoff $E_T \ (\Delta
E_T)$.  The legend of the curves are the same as in Fig. 2.

\item[{\rm Fig.\ 4.}] The signal cross section of Fig. 3(a,b,c)
satisfying a signal to background ratio $\geq 1$, are shown as functions
of $\tan\beta$ for $m_H = 80,100,120$ and 140  GeV by solid, dashed,
dot-dashed and dotted lines respectively.

\item[{\rm Fig. \ 5.}] The integrals of the signal and background
cross-sections are shown against cutoff $E_T \ (\Delta E_T)$ as in
Fig. 3, but for $\sqrt{s} = 4$ TeV.  The solid (dashed) and dot-dashed
(dotted) lines correspond to the signal (background) for $m_H = 100$
and 150 GeV respectively.  We take $\tan \beta = 1.4$.

\item[{\rm Fig. \ 6.}] The signal cross-sections of Fig. 5(a,b,c),
satisfying a signal to background ratio $\geq 1$, are shown as functions
of $\tan\beta$ for $m_H = 80,100,120,140$ and 150 GeV by solid,
dashed, dot-dashed, double-dot-dashed and dotted lines respectively.

\end{enumerate}


\begin{thebibliography}{999}
\bibitem{iso} The isolation cut strategy may be more appropriate at
the LHC, where the signal size is very large.
\end{thebibliography}
\end{document}